\documentclass[twocolumn,prb,superscriptaddress,longbibliography]{revtex4-2}

\usepackage{amsmath,amsfonts,amssymb}
\usepackage{graphicx,color}
\usepackage{hyperref}
\usepackage{epstopdf}
\usepackage{float}
\usepackage{multirow}
\usepackage{hhline}
\usepackage{ulem}
\usepackage{mathtools}
\usepackage[caption=false]{subfig}
\usepackage{soul}

\usepackage{graphicx}
\usepackage{dcolumn}
\usepackage{bm}

\usepackage[utf8]{inputenc} 

\usepackage{multirow}
\usepackage[warn]{mathtext}
\usepackage{amssymb}

\usepackage{textcomp} 
\usepackage{indentfirst} 
\usepackage{amsmath} 
\usepackage{graphicx}
\DeclareGraphicsExtensions{.pdf,.png,.jpg}
\usepackage{pgfplots}
\pgfplotsset{compat=1.13}

\usepackage{hyperref}
\hypersetup{
    colorlinks=true,
    linkcolor=blue,
    filecolor=black,      
    urlcolor=blue,
    citecolor= violet
}

\usepackage{algpseudocode}

\usepackage{adjustbox}
\usepackage{tabularx}
\usepackage[braket, qm]{qcircuit}

\usepackage{mathtools}

\usepackage{xcolor}
\definecolor{C0}{HTML}{4C72B0}
\definecolor{C1}{HTML}{DD8452}
\definecolor{C2}{HTML}{55A868}
\definecolor{C3}{HTML}{C44E52}

\usepackage{ulem}



\newcommand{\sectiontitle}[1]{\textbf{#1}}

\newcolumntype{Y}{>{\centering\arraybackslash}X}

\makeatletter
\def\@fnsymbol#1{\ensuremath{\ifcase#1\or \dagger\or \ddagger\or
   \mathsection\or \mathparagraph\or \|\or **\or \dagger\dagger
   \or \ddagger\ddagger \else\@ctrerr\fi}}
\makeatother

\footnotetext{These authors contributed equally to this work.}

\begin{document}

\title{Interleaved Benchmarking with Single-Qubit References}

\author{Ilya A. Simakov$^*$}
\email[Corresponding author: ]{simakov.ia@phystech.edu}
\affiliation{National University of Science and Technology ``MISIS'', 119049 Moscow, Russia}
\affiliation{Russian Quantum Center, 143025 Skolkovo, Moscow, Russia}

\author{Arina V. Zotova$^*$}
\affiliation{National University of Science and Technology ``MISIS'', 119049 Moscow, Russia}
\affiliation{Russian Quantum Center, 143025 Skolkovo, Moscow, Russia}
\affiliation{Moscow Institute of Physics and Technology, 141701 Dolgoprudny, Russia}

\author{Tatyana~A.~Chudakova}
\affiliation{National University of Science and Technology ``MISIS'', 119049 Moscow, Russia}
\affiliation{Russian Quantum Center, 143025 Skolkovo, Moscow, Russia}
\affiliation{Moscow Institute of Physics and Technology, 141701 Dolgoprudny, Russia}

\author{Alena~S.~Kazmina}
\affiliation{National University of Science and Technology ``MISIS'', 119049 Moscow, Russia}
\affiliation{Russian Quantum Center, 143025 Skolkovo, Moscow, Russia}

\author{Artyom~M.~Polyanskiy}
\affiliation{National University of Science and Technology ``MISIS'', 119049 Moscow, Russia}
\affiliation{Russian Quantum Center, 143025 Skolkovo, Moscow, Russia}
\affiliation{Moscow Institute of Physics and Technology, 141701 Dolgoprudny, Russia}

\author{Nikolay~N.~Abramov}
\affiliation{National University of Science and Technology ``MISIS'', 119049 Moscow, Russia}

\author{Mikhail~A.~Tarkhov}
\affiliation{Institute of Nanotechnology of Microelectronics, Russian Academy of Sciences, Moscow, 119991 Russia}

\author{Alexander~M.~Mumlyakov}
\affiliation{Institute of Nanotechnology of Microelectronics, Russian Academy of Sciences, Moscow, 119991 Russia}

\author{Igor~V.~Trofimov}
\affiliation{Institute of Nanotechnology of Microelectronics, Russian Academy of Sciences, Moscow, 119991 Russia}

\author{Nikita~Yu.~Rudenko}
\affiliation{National University of Science and Technology ``MISIS'', 119049 Moscow, Russia}

\author{Maxim~V.~Chichkov}
\affiliation{National University of Science and Technology ``MISIS'', 119049 Moscow, Russia}

\author{Vladimir~I.~Chichkov}
\affiliation{National University of Science and Technology ``MISIS'', 119049 Moscow, Russia}

\author{Grigoriy S. Mazhorin}
\affiliation{National University of Science and Technology ``MISIS'', 119049 Moscow, Russia}
\affiliation{Russian Quantum Center, 143025 Skolkovo, Moscow, Russia}

\date{\today}

\begin{abstract}

Cross-entropy benchmarking (XEB) with single-qubit reference sequences is widely used to characterize multi-qubit gates in large-scale quantum processors, despite the lack of a rigorous theoretical justification. Here we show that the commonly employed additive single-qubit errors approximation underlying this approach breaks down and leads to a systematic overestimation of gate fidelities. We derive an analytical expression for the joint decay of simultaneous single-qubit reference sequences, identify when interleaved circuits generate sufficient randomization for the standard depolarizing approximation, and obtain a refined expression for the interleaved gate fidelity estimation. Experiments on a superconducting quantum processor validate the theory and demonstrate that fidelities obtained using XEB with single-qubit references agree with those extracted from standard interleaved randomized benchmarking (IRB), while achieving higher precision due to reduced reference-sequence errors. Our results establish theoretical foundation for interleaved benchmarking protocols employing single-qubit reference sequences and show that, with appropriate post-processing, they provide a simple and accurate alternative to multi-qubit Clifford reference circuits.

\end{abstract}

\maketitle

\sectiontitle{Introduction.}
Interleaved randomized benchmarking (IRB) \cite{Magesan_2012} is the established standard for estimating the fidelity of individual quantum operations and it conventionally relies on reference circuits drawn from the multi-qubit Clifford group \cite{PracticalIntroduction, Randomizedbenchmarking}.
In experimental practice, however, the characterization of large-scale quantum devices is generally performed with cross-entropy benchmarking (XEB) employing single-qubit reference sequences \cite{Arute_2019, StrongQuantumComputationalAdvantage, PhysRevLett.129.030501, google2023suppressing, acharya2024quantumerrorcorrectionsurface, Modular_2024, gao2025establishing, chen2025efficient, RemoteSong, fan2025calibrating, wang2026demonstration}.
This approach substantially reduces experimental overhead by leveraging simple reference operations used for routine single-qubit calibration.
Moreover, XEB naturally supports non-Clifford target gates \cite{Foxen_2020, TunableInductiveCoupler, chen2025efficient, yang2026globalparametricgatesmultiqubit}, extends straightforwardly to multi-qubit operations beyond two qubits \cite{kim2022high}, and is well suited for error budget analysis \cite{Arute_2019, RealizingaContinuous_Zurich, mkadzik2025operating, simakov2023coupler, Threemode_tunable_coupler}.
Despite the widespread use and clear practical advantages, interleaved benchmarking protocols based on single-qubit references lacks a rigorous theoretical foundation.

benchmarking protocols employing single-qubit reference sequences lack a rigorous theoretical foundation.

Utilizing single-qubit references in interleaved XEB implicitly relies on two seemingly natural but unproven assumptions:
(i) the additivity of single-qubit errors in the joint fidelity decay, and (ii) the applicability of standard multi-qubit depolarizing approximation for interleaved circuits.
While both assumptions are supported by empirical evidence \cite{Foxen_2020}, neither has been formally confirmed.
If violated, these assumptions can lead to systematic biases in the extracted fidelities and obscure the identification of dominant error mechanisms in multi-qubit devices, crucial for gate-architecture improvement.

In this work, we develop a theoretical framework for interleaved benchmarking with single-qubit reference sequences. We demonstrate the breakdown of the commonly used additive approximation for joint fidelity decay and derive an analytical relation between the measured decay and individual-qubit depolarizing errors. 
We further show that, despite the non-exponential decay of simultaneous single-qubit benchmarking, interleaved sequences rapidly generate the effective randomization required for the multi-qubit depolarizing approximation, and introduce a practical criterion for identifying this regime.
Experiments on a superconducting quantum processor confirm the theoretical predictions and yield fidelity estimates consistent with standard IRB, while achieving substantially improved precision due to reduced reference-sequence errors.

\sectiontitle{Benchmarking preliminaries.}
The core idea behind gate benchmarking protocols such as XEB \cite{Arute_2019} and RB \cite{ScalableandRobustRandomizedBenchmarking} is to execute random gate sequences of varying length. Averaging over random circuits effectively twirls the underlying noise into a depolarizing channel. As a result, an initial state $\left| \psi \right\rangle$, evolved under random unitaries $U$ and averaged over the ensemble, takes the form
\begin{equation}
\overline{U^\dagger \rho_U U}=F \left| \psi \right\rangle \left\langle \psi \right|+(1 - F)I/d,
\label{eq:avg_U}
\end{equation}
with $I$ the identity operator, $d = 2^n$ the dimension of the $n$-qubit Hilbert space and $\rho_U$ are the density matrices resulting from the action of noisy sequences.
In this model, the state depolarizing fidelity $F$ averaged over random circuits of length $m$ exhibits an exponential decay
\begin{equation}
F = p^m,
\label{eq:power_law}
\end{equation}
where \(p\) denotes, throughout this work, the average depolarizing fidelity per gate block and is extracted directly from experimental data.

While this approach provides the average fidelity of a gate set, it does not quantify a specific operation. To address this issue, the interleaved  protocol was introduced. It consists of two stages: first, standard RB is performed to obtain the reference depolarizing fidelity $p_{\text{ref}}$; second, the experiment is repeated with the target gate $G$ interleaved between random Cliffords, yielding $p_{\text{int}}$. The ratio $p_G = p_{\text{int}}/p_{\text{ref}}$ gives the depolarizing fidelity of the particular operation.

The depolarizing fidelity $p$ is then related to the conventional average gate fidelity~\cite{nielsen2002simple, pedersen2007fidelity, PhysRevLett109060501} via
\begin{equation}
\mathcal{F}=\frac{\mathrm{Tr}(E) + d}{d(d+1)}=\frac{d-1}{d} p + \frac{1}{d}.
\label{eq:conventional_fidelity}
\end{equation}
Here, \(E\) is the Pauli transfer matrix of the error channel, defined through the relation \(R = E R_{\mathrm{ideal}}\), where \(R\) and \(R_{\mathrm{ideal}}\) denote the implemented and ideal gates, respectively. Together, these relations constitute the theoretical framework of randomized and cross-entropy benchmarking and form the basis for quantitative characterization of quantum processors.

\sectiontitle{Reference sequences under local noise.}
The goal of the present work is to extend the standard benchmarking framework to enable a more experimentally convenient averaging strategy, directly relating the multi-qubit decay to individual-qubit errors. In modern quantum processors, gate performance is largely constrained by decoherence mechanisms, such as dephasing and energy relaxation, which act on distinct qubits. Therefore, in addition to standard RB noise requirements, we assume a local noise model, composed of independent single-qubit error channels, which accurately captures the dominant error sources in contemporary devices. Within this setting, we examine two averaging approaches widely used in experimental practice: averaging over multi-qubit Clifford group, and averaging over simultaneously executed single-qubit gates $\mathcal{C}_1^{\otimes n}$, yielding the standard multi-qubit and single-qubit depolarizing models, respectively.

We begin with the standard benchmarking protocol, in which reference circuits are composed of $n$-qubit Clifford operations forming a unitary 2-design~\cite{Gross_2007}.
In this case, the average fidelity is determined via Eq.~\eqref{eq:conventional_fidelity}, with the error channel taking the product form \(E = \bigotimes_{i=1}^n \operatorname{diag}(1, p_i, p_i, p_i),\)
where the diagonal representation is given in the $\{I,X,Y,Z\}$ Pauli basis and $p_i$ denotes depolarizing fidelity associated with the $i$-th qubit.
To leading order in the individual error rates $e_i = 1 - p_i$, the resulting joint decay of the depolarizing fidelity is
\begin{equation}
F_{\mathrm{multi}} = \left( p_{\mathrm{multi}} \right)^m
= \left( 1 - \frac{3}{4} \cdot \frac{1}{1 - (1/4)^n} \sum_{i=1}^n e_i \right)^m .
\label{eq:error_clif}
\end{equation}
Thus, the exponential decay law is preserved, with an effective depolarization fidelity per gate block $p_{\mathrm{multi}}$.

Next, we consider simultaneous single-qubit benchmarking, a protocol used to assess single-qubit gate performance in multi-qubit systems. The fidelity measured in the benchmarking procedure is defined as the difference between the experimentally extracted and ideal bitstring probabilities of the final states. In Appendix~\ref{app:fidelity_decay}, we analyze these outcome probabilities for a single qubit and extend the analysis to $n$ independent qubits using a combinatorial method. Specifically, we replace averaging over random unitaries $U$ with an explicit summation over all possible measurement outcomes generated solely by single-qubit Clifford operations. The resulting expression for the average depolarizing fidelity of the final state as a function of the sequence length $m$ is
\begin{equation}
F_{\mathrm{single}} =\frac{
2^n \prod_{i=1}^n \left( 2 + p_i^m \right)+ 3^n
- \prod_{i=1}^n \left( 3 + p_i^m \right)- 4^n}
{6^n + 3^n - 2 \cdot 4^n}.
\label{eq:error_sq}
\end{equation}
In contrast to the multi-qubit Clifford case, this expression exhibits a pronounced deviation from a simple exponential decay.

\begin{figure}[t]
    \centering
    \includegraphics[width=\linewidth]{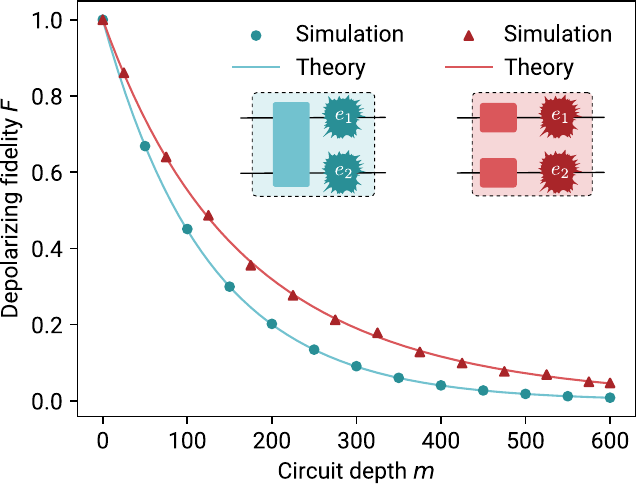}
    \caption{Simulation of the XEB experiment using the same physical relaxation and dephasing noise model for both protocols. Teal circles correspond to two-qubit Clifford gates, and red triangles to simultaneous single-qubit Cliffords. The solid curves show the theoretical predictions $F_\text{multi}$~[Eq.~(\ref{eq:error_clif})] and $F_\text{single}$~[Eq.~(\ref{eq:error_sq})]  with the equivalent depolarizing error rates.}
    \label{fig:models}
\end{figure}

To validate the analytical results and directly compare the two averaging approaches, we perform density-matrix simulations of standard two-qubit and simultaneous single-qubit XEB experiments using an identical physical noise model for both protocols, with relaxation and dephasing parameters representative of our superconducting device~\cite{mazhorin2026nativecczgatefluxonium}. The corresponding quantum channels are implemented through Kraus maps applied after each gate. The results are shown in Fig.~\ref{fig:models} (circles and triangles). The corresponding effective depolarizing error rates are calculated from the physical noise channels using the standard first-order relation~\cite{nielsen2010quantum, pedersen2007fidelity} $e=(t/T_1+t/T_\varphi)/3$, where $t$ denotes the gate duration, yielding $e_1 = 0.006$ and $e_2 = 0.004$. The theoretical predictions $F_\text{multi}$ and $F_\text{single}$ obtained with these error rates are in excellent agreement with the numerical results. Despite being simulated with the same noise model, the two protocols exhibit markedly different decays, demonstrating the impact of the circuit averaging procedure.

\begin{figure}[t]
    \centering
    \includegraphics[width=\linewidth]{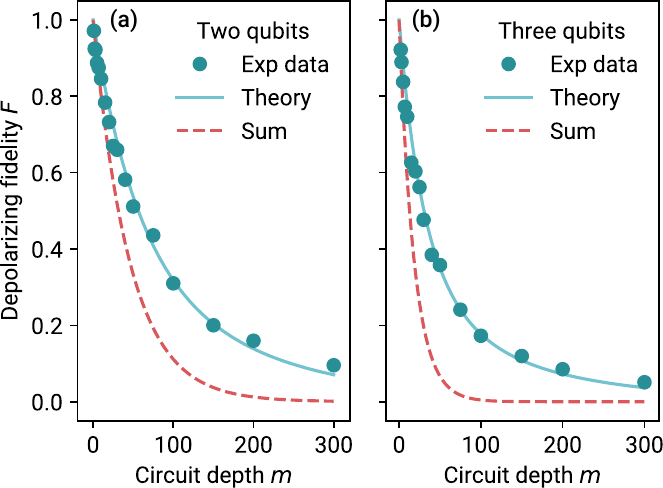}
    \caption{Simultaneous single-qubit cross-entropy benchmarking for (a) two and (b) three qubits. Teal points show the experimentally measured fidelities, error bars are obtained using a bootstrap technique. The teal solid curve represents the theoretical dependence calculated via Eq.~(\ref{eq:error_sq}) using depolarizing errors $e_i$ extracted from individual single-qubit benchmarking. The dashed red curve corresponds to the additive model $F=(1-\sum_i e_i)^m$, illustrating the substantial deviation from the experimentally observed decay.}
    \label{fig:simultaneous_1q_xeb}
\end{figure}

We further verify Eq.~\eqref{eq:error_sq} experimentally by performing simultaneous single-qubit benchmarking on two- and three-qubit subsets of a superconducting fluxonium-based processor~\cite{mazhorin2026nativecczgatefluxonium}. The measured decays (teal dots in Fig.~\ref{fig:simultaneous_1q_xeb}) are well described by the theoretical model (teal curve), constructed using individual depolarizing errors obtained from isolated single-qubit benchmarking experiments.

These results therefore demonstrate that the choice of averaging procedure has a substantial impact on the observed fidelity decay. In particular, the commonly used additive approximation $F = (1 - \sum_i e_i)^m$, shown with a dashed red curve in Fig.~\ref{fig:simultaneous_1q_xeb}, fails to capture the joint behavior in simultaneous single-qubit benchmarking. At the same time, both averaging approaches remain fundamentally linked through the same set of individual depolarizing errors.

\begin{figure}[t]
    \centering
    \includegraphics[width=\linewidth]{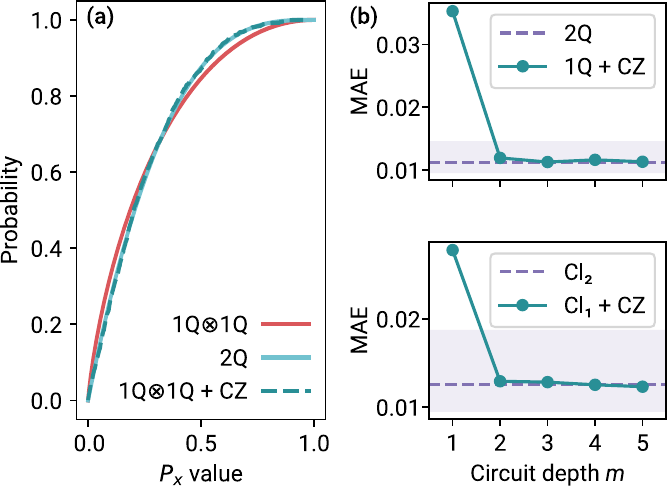}
    \caption{(a) Cumulative distribution functions of output probabilities for the limiting ensembles of arbitrary two-qubit and factorized single-qubit random circuits, together with depth-4 interleaved CZ sequences. (b) Mean absolute error (MAE) relative to the corresponding limiting ensemble versus the number of interleaved CZ layers. The shaded region shows the mean $\pm$ one standard deviation for an ideal random generator (100 samples). In both arbitrary unitary and Clifford twirling, the scrambling criterion is satisfied after two interleaved layers.    }
    \label{fig:PT}
\end{figure}

\sectiontitle{Scrambling criterion for interleaved circuits.}
The non-exponential joint decay established above raises the question of whether interleaved benchmarking circuits with single-qubit references generates sufficient effective twirling of errors to justify the use of the standard depolarizing model.
This requires interleaved circuits to produce the same effective randomization as a fully chaotic ensemble uniformly sampling the $n$-qubit Hilbert space. When the condition is satisfied, the measurement outcome probabilities follow the Porter--Thomas (PT) distribution~\cite{PhysRev.104.483},
\begin{equation}
\mathcal{P}_{\mathrm{PT}}(P_x)
=
(d-1)(1-P_x)^{d-2},
\label{eq:PT_density}
\end{equation}
where $P_x$ is the probability of measuring bitstring $x$.
By themselves, factorized single-qubit random circuits produce a different distribution,
\color{black}
\begin{equation}
\mathcal{P}_{\mathrm{1Q}}(P_x)
=
\frac{(-\ln P_x)^{n-1}}{(n-1)!},
\label{eq:AN_density}
\end{equation}
derived in Appendix~\ref{app:1qdistr}. This results in error twirling into independent single-qubit depolarizing channels rather than a multi-qubit one. Fig.~\ref{fig:PT}(a) shows the corresponding cumulative distribution functions for the two-qubit case.

Sufficient scrambling is achieved when the output statistics of an interleaved benchmarking sequence become indistinguishable from the PT distribution (or, equivalently for benchmarking purposes, from those generated by random $n$-qubit Clifford circuits). To quantify this agreement, we use the mean absolute error (MAE) between the sampled cumulative distribution function $\mathcal{C}_{\mathrm{samp}}$ and the limiting distribution $\mathcal{C}_{\mathrm{lim}}$,
\begin{equation}
\mathrm{MAE}(\mathcal{C}_{\mathrm{samp}},\mathcal{C}_{\mathrm{lim}})
=
\frac{1}{M}
\sum_{i=1}^{M}
\left|
\mathcal{C}_{\mathrm{samp}}(p_i)
-
\mathcal{C}_{\mathrm{lim}}(p_i)
\right|,
\end{equation}
where $p_i$ are the output probabilities and $M$ is the total number of sampled probabilities. Since a benchmarking experiment samples only a finite number of random circuits, even an ideal random generator does not exactly reproduce the limiting distribution. Consequently, the expected MAE remains finite even for an ideal random generator and depends on the number of sampled circuits. Thus, we formulate the scrambling criterion as follows: An interleaved protocol generates sufficient scrambling when its average MAE lies within one standard deviation of the average MAE obtained for an ideal random generator with the same number of sampled circuits.

Fig.~\ref{fig:PT}(b) illustrates this procedure for an interleaved benchmarking of a CZ gate using arbitrary unitary and Clifford single-qubit twirling. In both cases, the criterion is satisfied after only two interleaved layers. Applying the same analysis to widely used entangling gates, including iSWAP, MS, B~\cite{chen2025efficient}, and CCZ, shows that they also meet the same criterion after only a few circuit layers (Appendix~\ref{app:scrambling}). These results demonstrate that for a broad class of entangling gates interleaved sequences with single-qubit references rapidly produce the effective randomization required for the depolarizing approximation.

\sectiontitle{Interleaved gate fidelity extraction.}
Once the scrambling criterion is satisfied, Eq.~\eqref{eq:error_clif} provides the effective reference depolarization required to extract the interleaved gate fidelity using the standard relation. 
For the two-qubit case, it becomes
\begin{equation}
    p_G
    =
    \frac{p_{\mathrm{int}}}{p_{\mathrm{multi}}}
    =
    \frac{p_{\mathrm{int}}}
    {1-\frac{4}{5}(e_1+e_2)}.
    \label{eq:int_1Q_sim}
\end{equation}

\begin{figure}[t]
    \centering
    \includegraphics[width=\linewidth]{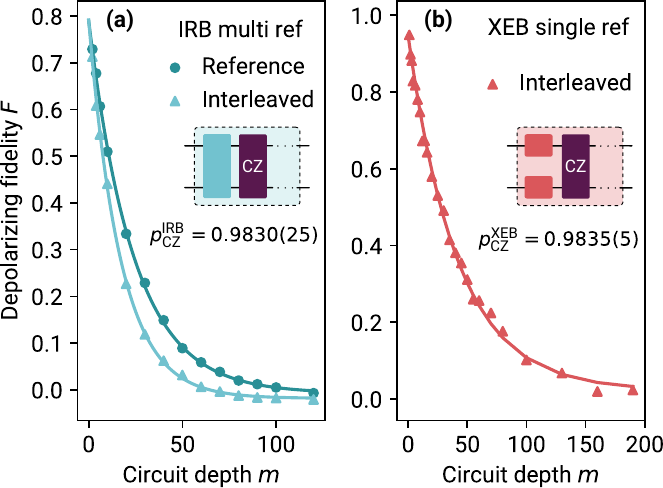}
    \caption{Comparison of Clifford-based standard interleaved randomized benchmarking (IRB) with a multi-qubit reference sequence and cross-entropy benchmarking (XEB) with a single-qubit reference. The interleaved CZ gate fidelity calculated using the refined formula~(\ref{eq:int_1Q_sim}) is consistent with that obtained from the IRB experiment, while achieving higher accuracy due to reduced errors in the reference sequence. Uncertainties are estimated using bootstrap resampling.}
    \label{fig:interleaved}
\end{figure}

We experimentally validate Eq.~\eqref{eq:int_1Q_sim} on the fluxonium-based quantum processor~\cite{mazhorin2026nativecczgatefluxonium} 
by comparing standard IRB with interleaved XEB using single-qubit references (Fig.~\ref{fig:interleaved}). IRB, averaged over 160 random circuits, yields a depolarizing gate fidelity of $p_{\mathrm{CZ}}^{\mathrm{multi}}=0.9830(25)$.
The decay parameters $p_{\mathrm{int}}$ and $p_{\mathrm{ref}}$ are obtained by fitting the measured fidelities to the exponential form $Ap^m$, where the prefactor $A$ accounts for state preparation and measurement errors. 
XEB, averaged over 50 random circuits, gives $p_{\mathrm{CZ}}^{\mathrm{single}}=0.9835(5)$ using Eq.~\eqref{eq:int_1Q_sim}, where the single-qubit depolarizing errors $e_1$ and $e_2$ are measured individually in reference XEB experiment. The two values agree within their uncertainties, experimentally validating the proposed extraction procedure.

More generally, the proposed framework applies to arbitrary system size. Neglecting the averaging correction leads to a systematic overestimation of the target-gate fidelity, with the resulting bias increasing with the number of qubits and approaching $\sum_i e_i/4$ to leading order. In the present experiment, this would give $p_{\mathrm{CZ}}=0.9850(5)$ instead of $0.9835(5)$.

\color{black}

\sectiontitle{Discussion.}
Employing single-qubit references not only yields a correct estimate of the interleaved gate fidelity, but also provides higher precision than standard interleaved randomized benchmarking. Because single-qubit gates are substantially more accurate than multi-qubit operations, the reference and interleaved sequences decay much more slowly than circuits built from $n$-qubit Cliffords. Consequently, the decay  $F(m)\sim p^m$ can be fitted over a wider range of circuit depths $\sim 1/(1-p)$, improving the precision of the extracted depolarizing fidelity and, as a result, the target-gate fidelity.

Neither the scrambling criterion nor the corrected noise-averaging procedure are specific to XEB. More generally, they apply to benchmarking protocols that characterize multi-qubit operations using single-qubit references. In particular, cyclic randomized benchmarking (CRB)~\cite{Erhard_2019} often employs single-qubit reference fidelities because they enable the fidelity of a “bare” multi-qubit gate to be extracted with minimal experimental overhead~\cite{PracticalIntroduction, Mitchell_2021, kim2022high}. In such cases, the same conversion from single-qubit depolarization to the corresponding multi-qubit depolarization is required to avoid overestimation of the target-gate fidelity.

A separate issue concerns the presence of nonlocal errors, which can lead the reference-sequence decay to deviate from the form $F_{\mathrm{single}}$.
A practical experimental consistency check for the local-noise model applicability is the agreement between single-qubit fidelities extracted from isolated and simultaneous benchmarking, since correlated errors generally contribute only in the latter case \cite{Gambetta_2012}.
Whether nonlocal errors can be reliably detected and quantitatively characterized using simultaneous single-qubit benchmarking alone remains an open question and an important direction for future research.

\sectiontitle{Conclusion.}
This work provides the theoretical foundation for interleaved benchmarking with single-qubit reference sequences by identifying and resolving the two key assumptions underlying the method. We show that the additivity of single-qubit errors in the joint reference-sequence decay is generally invalid, derive the correct analytical relation between the joint decay and the individual single-qubit depolarizing fidelities, and validate it using numerical simulations and experiments on a superconducting quantum processor. Next, we formulate a practical criterion for determining when repeated interleaved layers generate global randomization. 
Applying this criterion, we show that the interleaved sequences for widely used two- and multi-qubit entangling gates follow the standard depolarizing description, justifying the use of single-qubit twirling for their characterization.

Together, these results establish a simple operational procedure for high-precision estimation of multi-qubit gate fidelities. Once the scrambling criterion is satisfied, the measured single-qubit depolarizing errors are converted into the corresponding effective multi-qubit depolarization, which is then used to extract the target-gate fidelity. The framework is therefore not specific to XEB, but applies generally to benchmarking protocols employing simplified references. As quantum processors scale up, this technique provides a practical and theoretically well-founded approach for accurate and high-precision characterization of multi-qubit quantum gates.
\newline

\textit{Acknowledgments.} 
We thank Alexey Ustinov and Ilya Besedin for helpful discussions and critical comments on the manuscript. 
The work was supported by Rosatom in the framework of the Roadmap for Quantum computing (Contract No. 868/1726-D dated September 22, 2025)
and by the Ministry of Science and Higher Education of the Russian Federation in the framework of the Program of Strategic Academic Leadership “Priority 2030” (MISIS Strategic Technology Project Quantum Internet).\newline

\appendix

\section{Fidelity decay for a simultaneous single-qubit benchmarking}
\label{app:fidelity_decay}

To derive Eq.~(\ref{eq:error_sq}), we consider cross-entropy benchmarking from an experimental perspective, following the notations of tutorial~\cite{GoogleXEBwebpage}. In XEB, the experimentally measured output distribution of bitstrings $x$ is compared with the ideal one computed on classical computer. Taking the componentwise product of the bitstring probabilities in each term of Eq.~(\ref{eq:avg_U}) with the ideal probabilities $p_\text{ideal}(x)$ one gets
\begin{equation}
    \begin{split}
        m_U &= \sum_x p_\text{ideal}(x) p_\text{meas}(x), \\
        e_U &= \sum_x p_\text{ideal}(x) p_\text{ideal}(x), \\
        u_U &= \frac{1}{2^N} \sum_x p_\text{ideal}(x) = \frac{1}{2^n}.
    \end{split}
\end{equation}
Here $p_\text{meas}(x)$ is the experimentally measured probability of obtaining the bitstring $x$. Within the depolarizing model, the measured probabilities are assumed to follow a mixture of the ideal distribution and the uniform distribution, which leads to the linear relation
\begin{equation}
    m_U = F e_U + (1-F) u_U.
\end{equation}
The state depolarizing fidelity $F$ is then extracted using the least squares method, yielding
\begin{equation}
    F = \frac{\sum_U (m_U - u_U) (e_U - u_U)}{\sum_U (e_U - u_U)^2}.
    \label{eq:f_google_mu_uu}
\end{equation}

We first analyze the single-qubit case. Since the benchmarking sequences consist of Clifford operations, the ideal final state is aligned with the $z$ axis of the Bloch sphere with probability $1/3$ and lies in the $xy$ plane with probability $2/3$.
Assuming that the qubit depolarizes with probability $\delta = 1 - p^m$, if the ideal final state is aligned with $z$ axis,
$m_U = 1 - \delta/2$ and $e_U = 1,$
whereas for states in the $xy$ plane,
$m_U = e_U = u_U = 1/2$.
Hence, only realizations when the final state is aligned with $z$ give a nonzero contribution to Eq.~(\ref{eq:f_google_mu_uu}).

For $n$ qubits, we have to account for all possible configurations in which $k$ qubits end up aligned along the $z$ axis. Since these events occur independently for each qubit, the averaging over random circuits $U$ can be replaced by a sum over $k$. As a result, the denominator of Eq.~(\ref{eq:f_google_mu_uu}) becomes
\begin{equation}
    \sum_U (e_U - u_U)^2 = \sum_{k=0}^n C_n^k \left( \frac{1}{3} \right)^k \left( \frac{2}{3} \right)^{n-k} 
    \left( \frac{1}{2^{n-k}} - \frac{1}{2^n} \right)^2,
\end{equation}
where $C_n^k$ denotes the binomial coefficient.

The numerator depends on both the ideal and the measured probabilities. 
Because the depolarization fidelities $\delta_i$ are qubit dependent, the summation over the $k$ selected qubits cannot be reduced to a simple combinatorial factor and is carried out explicitly over all $k$-qubit subsets $\{i_1,\dots,i_k\}$. The numerator of Eq.~(\ref{eq:f_google_mu_uu}) then reads
\begin{equation}
\begin{split}
&\sum_U (m_U - u_U)(e_U - u_U) =
\sum_{k=0}^n
\left( \frac{1}{3} \right)^k
\left( \frac{2}{3} \right)^{n-k}\times \\
& \sum_{1 \leq i_1 < \dots < i_{k} \leq n}
\biggl[
\frac{1}{2^{n-k}}
\prod_{j=1}^{k}\left(1-\frac{\delta_{i_j}}{2}\right)
- \frac{1}{2^n}
\biggr]
\left[ \frac{1}{2^{n-k}} - \frac{1}{2^n} \right].
\end{split}
\end{equation}

By expanding the brackets and applying the generating-function decomposition
\begin{equation}
     \prod_{i=1}^n \left( 1+a_ix \right) = \sum_{k=0}^n \ \sum_{1 \leq i_1 < ... < i_{k} \leq n} a_{i_1} a_{i_2} \cdot\cdot\cdot a_{i_k} x^k
\end{equation}
and, as its special case, the binomial expansion $\sum_{k=0}^n C_n^k a^k b^{n-k} = (a+b)^n$, the fraction~(\ref{eq:f_google_mu_uu}) results in
\begin{equation} 
F = \frac{2^n \prod_{i=1}^n \left( 3-\delta_i \right) + 3^n - \prod_{i=1}^n \left( 4-\delta_i \right) - 4^n}{6^n + 3^n - 2\cdot4^n},
\end{equation} 
which, after substitution $\delta_i = 1 - p_i^m$, yields Eq.~(\ref{eq:error_sq}).

\section{Factorized single-qubit outcome bitstring distribution} 
\label{app:1qdistr}

The distinction between simultaneous single-qubit and multi-qubit reference benchmarking can be quantified through the probability density of measured bitstring outcomes. For a fully chaotic quantum state uniformly sampled over the $n$-qubit Hilbert space, these probabilities follow the Porter-Thomas distribution~(\ref{eq:PT_density}). In contrast, for factorized random circuit outcome bitstrings, or equivalently for the diagonal elements of the outcome density matrix, the resulting distribution differs from the Porter-Thomas form.

For two qubits,
\begin{equation}
\ket{\psi} = (a_0\ket{0} + a_1\ket{1}) \otimes (b_0\ket{0} + b_1\ket{1}),
\end{equation}
the outcome bitstring $ij$ probabilities are
\begin{equation}
P_{ij} = |a_i|^2 |b_j|^2 = P_i^{(1)} P_j^{(2)}, \quad i,j \in \{0,1\}.
\end{equation}
Here, $P_i^{(1)}$ and $P_j^{(2)}$ are independent random variables. Therefore, the probability density function of their product can be obtained using the standard expression for the product $Z = X \cdot Y$ of independent variables $X$ and $Y$:
\begin{equation}
	f_{Z}(z) = \int_{-\infty}^{\infty} f_{X}(x) f_{Y}\left(\frac{z}{x}\right) \frac{1}{|x|} \, dx.
	\label{eq:product}
\end{equation}
Substituting the single-qubit Porter-Thomas distribution and restricting the integration limits to $[x=z, \, 1]$, one finds
\begin{equation}
	\mathcal{P}_{1Q}(P_x) = \int_{P_x}^1 \frac{1}{x} dx = -\ln P_x, \quad \text{for} \; N=2
\end{equation}
For $n$ independent qubits applying Eq.~(\ref{eq:product}) recursively $(n-1)$ times gives in Eq.~(\ref{eq:AN_density}) of the main text.

Finally, when random circuits are composed exclusively of Clifford operations, the diagonal elements of the resulting density matrix take only discrete values $2^k/2^n$, where $k = 0, \ldots, n$. Consequently, the resulting probability distribution function acquires a step-like profile, with the step heights determined by the relative frequency of these diagonal elements across all states that can be generated within the Clifford group, both in the single- and multi-qubit cases.

\begin{figure}[b]
    \centering
    \includegraphics[width=\linewidth]{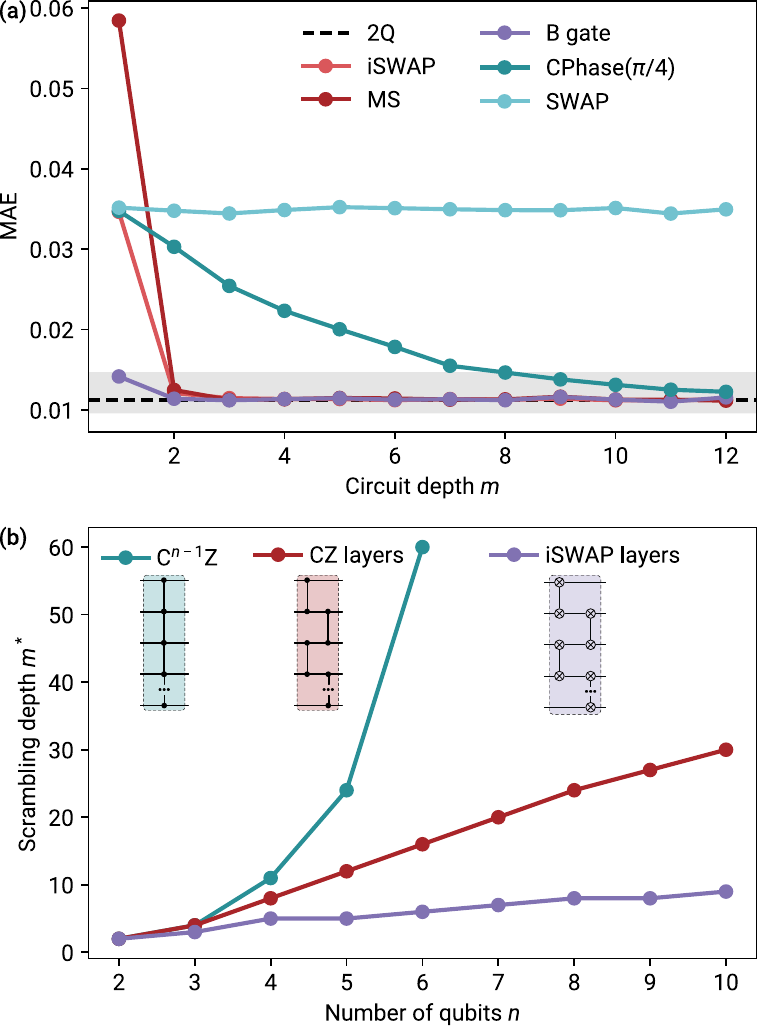}
    \caption{(a) MAE between the output-state statistics of interleaved circuits and the cumulative Porter-Thomas distribution for different interleaved gates. The shaded region represents the mean $\pm$ standard deviation of an ideal random ensemble. Each distribution is estimated from 100 randomly sampled circuits. (b) Scrambling depth $m^*$ versus system size for $\mathrm{C}^{n-1}\mathrm{Z}$, nearest-neighbor CZ and iSWAP layer structures.}
    \label{fig:different_gates}
\end{figure}

\section{Applicability to different gates}
\label{app:scrambling}

Here, we apply the scrambling criterion introduced in the main text to a range of widely used quantum gates. Fig.~\ref{fig:different_gates}(a) shows the MAE between the output-state statistics of different interleaved circuits and the cumulative Porter-Thomas distribution. Commonly used entangling gates employed in universal quantum computing, such as CZ, iSWAP, B, and MS, satisfy the criterion after only two interleaved layers. By contrast, CPhase$(\pi/4)$ requires substantially greater depth, while SWAP never satisfies the criterion because the resulting ensemble remains factorized. 
These examples illustrate both the broad scope of the proposed framework and its ability to identify gates for which the scrambling assumption does not hold.

Fig.~\ref{fig:different_gates}(b) shows how the scrambling depth $m^*$, defined as the smallest depth at which the interleaved circuit becomes statistically consistent with the ideal random generator, scales with system size for three layer structures: multi-controlled $\mathrm{C}^{n-1}\mathrm{Z}$, nearest-neighbor CZ, and nearest-neighbor iSWAP layers. In all three cases, the scrambling criterion is satisfied, although the required scrambling depth increases with system size at a gate-dependent rate. Nevertheless, this overhead remains substantially smaller than that associated with constructing multi-qubit Clifford references, preserving the practical advantage of the proposed framework for larger systems.

\normalem{}
\bibliography{main}

\end{document}